# Classification and Region Analysis of COVID-19 Infection using Lung CT Images and Deep Convolutional Neural Networks

Saddam Hussain Khan[1], Anabia Sohail[1], Asifullah Khan[1, 2]*, Yeon Soo Lee[3]

asif@pieas.edu.pk

[1]Pattern Recognition Lab, Department of Computer & Information Sciences, Pakistan Institute of Engineering & Applied Sciences, Nilore, Islamabad 45650, Pakistan,

[2]PIEAS Artificial Intelligence Center (PAIC), Pakistan Institute of Engineering & Applied Sciences, Nilore, Islamabad 45650, Pakistan

[3]Deparment of Biomedical Engineering, College of Medical Sciences, Catholic University of Daegu, South Korea

## Abstract

COVID-19 is a global health problem. Consequently, early detection and analysis of the infection patterns are crucial for controlling infection spread as well as devising a treatment plan. This work proposes a two-stage deep Convolutional Neural Networks (CNNs) based framework for delineation of COVID-19 infected regions in Lung CT images. In the first stage, initially, COVID-19 specific CT image features are enhanced using a two-level discrete wavelet transformation. These enhanced CT images are then classified using the proposed custom-made deep CoV-CTNet. In the second stage, the CT images classified as infectious images are provided to the segmentation models for the identification and analysis of COVID-19 infectious regions. In this regard, we propose a novel semantic segmentation model CoV-RASeg, which systematically uses average and max pooling operations in the encoder and decoder blocks. This systematic utilization of max and average pooling operations helps the proposed CoV-RASeg in simultaneously learning both the boundaries and region homogeneity. Moreover, the idea of attention is incorporated to deal with mildly infected regions. The proposed two-stage framework is evaluated on a standard Lung CT image dataset, and its performance is compared with the existing deep CNN models. The performance of the proposed CoV-CTNet is evaluated using Mathew Correlation Coefficient (MCC) measure (0.98) and that of proposed CoV-RASeg using Dice Similarity (DS) score (0.95). The promising results on an unseen test set suggest that the proposed framework has the potential to help the radiologists in the identification and analysis of COVID-19 infected regions.

# 1. Introduction

COVID-19 is a contagious disease that primarily originated in Wuhan province of China in December 2019; however, in early 2020 it spread across the world [1], and to date, it persists with its devastating effects across all continents [2]. COVID-19 is most commonly manifested by mild flue like symptoms such as cough, fever, myalgia, and fatigue [3]. However, in severe cases, it causes alveolar damage, pneumonia, sepsis, and respiratory failure, which eventually lead to death [4].

The commonly used tests for assessment of COVID-19 patients are gene sequencing, reverse transcription polymerase chain reaction (RT-PCR), X-Ray and computed tomography (CT) based imaging techniques [5], [6]. Out of these aforementioned assays, RT-PCR and gene sequencing are considered as a gold standard. However, because these standard assays are expensive, many developing countries have a limited number of testing kits or lack the sequencing facilities. RT-PCR usually takes up to 2 days for detection and often suffers from the inherited limitation of viral RNA instability. Thus, it requires serial testing to eliminate the likelihood of false-negative results (RT-PCR detection rate is ~ 30% to 60%) and necessitates some additional supplementary tests [7]–[9]. In this regard, additional detection methods with high precision are also required for immediate treatment of the infected persons and to cease the transmission of contagious COVID-19 infectious.

CT based imaging is not expensive and available in all the hospitals and reliable, practical, and efficient tools for detection, prognosis and follow-up of COVID-19 patients. Several examination setups have shown reliable diagnostic power of CT imaging in capturing the lung abnormalities for COVID-19 infected individuals even when characteristic clinical symptoms are imperceptible, and RT–PCR is reported as a false negative [10], [11]. The characteristic radiographic features for the infected patients are bilateral patchy shadows, ground glass pacification (GGO), consolidation, pleural effusion, rounded morphology, and peripheral lung distribution [4], [12].

In a public health emergency, especially in epidemic and pandemics, there is a significant burden on hospitals and physicians. Visual analysis of a large number of CT images is a huge burden on radiologists. There are many areas where there are no experienced radiologists. The increased workload on radiologists affects performance. Moreover, radiologists are less sensitive and more specific towards identifying COVID-19 infection by analyzing CT images. In this regard, there is an urgent need for an automated tool that can aid the radiologists to improve performance and to deal with a large number of patients. Deep learning (DL) based diagnostic systems are very valuable tools in plunging the physicians. Previously, DL based automated systems has been employed to facilitate the radiologists in the identification of lung-related anomalies [13], [14]. The advantage of such a system is that they are reproducible and can sense the detect minute irregularities that cannot be located through a visual examination. In this ongoing COVID-19 pandemic era, several research groups have paid attention to develop automated systems to identify the COVID-19 infected individual by examining CT images [15]–[17].

In this work, we proposed a classification and segmentation based deep Convolutional Neural Network (CNN) framework to identify COVID-19 infected samples and to analyze their extant of spread and infection pattern on lungs CT scans. In the proposed workflow, we initially screen the CT samples for COVID-19 infection using proposed deep CNN classifier CoV-CTNet (also known as PIEAS Classification Network-1 (PCNet-1)). Whereas in the next stage, COVID-19 classified CT samples are further analyzed for identifying the lung regions exhibiting infection using proposed region approximation based semantic segmentation architecture CoV-RASeg (also known as PIEAS Segmentation Network-1 (PSNet-1)). The proposed framework is assessed on standard publicly available lung CT dataset, and its performance is compared against well-known existing deep architectures. Key contributions of this study are:

1. A two-stage framework consisting of classification and segmentation models is proposed for detection and region analysis of COVID-19 infected regions in the lungs.
2. A custom-made deep CNN based classification network is proposed that can effectively learn the CT image features of COVID-19 infection.
3. A novel deep CNN based semantic segmentation architecture is proposed for fine-scale demarcation of lung regions infected from COVID-19. For this purpose, max and average pooling operations are incorporated systematically in each encoder and decoder block.
4. Performance of the proposed deep CNNs based classification and segmentation models is compared with different existing models that are trained from scratch as well as fine-tuned using TL. Moreover, the idea of attention is incorporated in deep segmentation models to deal with sparse representation of infected regions and for effective segmentation of mildly infected lung regions.

The layout of the paper is as follows: Section 2 gives an insight into COVID-19 related work. Section 3 explains the Methodology of the proposed framework, whereas Section 4 presents the implementation details. Section 5 analyzes the results and discusses the performance of the implemented experiments, and finally, section 6 makes a conclusion.

## 2. Related Works

Nowadays, CT technology has been used for the analysis of COVID-19 disease in multiple countries like China, Spain, and Italy. However, CT scan analysis is usually tedious, time-consuming, and prone to human error. Therefore, machine learning (ML) based diagnostic tools are designed for fast and accurate image analysis as well as facilitate the medical staff. AI and Deep Learning (DL) models have already shown impressive performance in the medical field [18]. Especially, the deep CNN has become the workhorse for image classification, detection, localization, and segmentation. Several pre-trained CNN models with different approaches have been employed for the analysis of COVID-19 infected X-ray and CT images. Likewise, different researchers exploited the potential of TL for the prediction of COVID-19 infected image. In this regard, diverse TL-based fine-tuned CNN models like AlexNet, VGG, GoogleNet, ResNet, Dense Net, etc. have been evaluated on COVID-19 infected CT images, and their performance accuracy varies from 87% to 98% [19]–[21]. The COVID-Net was inspired by ResNet and was used to differentiate multi types of COVID-19 infections from normal pneumonia. Although COVID-Net has good accuracy (92%), yet it has a low detection rate (87%) [22]. Similarly, COVID-CAPS inspired by Capsule Net also reported good accuracy (98%), but it is less sensitive (80%) towards COVID-19 infection [23]. Moreover, a novel classification model COVID-RENet inspired by smooth and boundary information of images and achieved 98% accuracy. All these pre-trained CNN models have been trained on Natural images and fine-tuned on the COVID-19 dataset [24].

On the other hand, segmentation of the infected regions is usually performed to identify the location of disease and severity. Initially, some classical segmentation techniques like watershed have been employed but fail to show satisfactory performance [25], [26]. Therefore,

DL based 'VB-Net' has been introduced for the segmentation of COVID-19 infected regions using CT images and reported the dice similarity (DS) score of 91%. Moreover, the COVID-19 JCS system based on classification and segmentation has been developed to visualize and segment the infected region. The JCS system reported the 95.0% detection rate, 93.0% specificity on classification, while low DS score (78.3%) on segmentation. However, most of the existing COVID-19 diagnostic systems have been trained on a small amount of CT datasets. Usually, these diagnostic systems have mostly two main challenges: 1) unavailability of sufficient amount of training data, which is required to make deep CNN models robust towards diverse categories of COVID-19 infections; 2) detection is restricted to the classification of infected samples and lacks the information of the infected region and severity of the disease.

## 3. Methodology

In this study, we proposed classification and segmentation based deep CNN framework for automatic analysis of COVID-19 abnormalities in lung CT images. The proposed framework is constituted of classification and segmentation stages. In the first stage, COVID-19 infected individuals are segregated from healthy CT samples by performing classification. Whereas, in the second stage, the segmentation of COVID-19 infected regions on lung CT images is performed to obtain fine-scale region details. The segmentation of infected regions can be helpful in quantifying infection spread. The detailed workflow is shown in Figure 1.

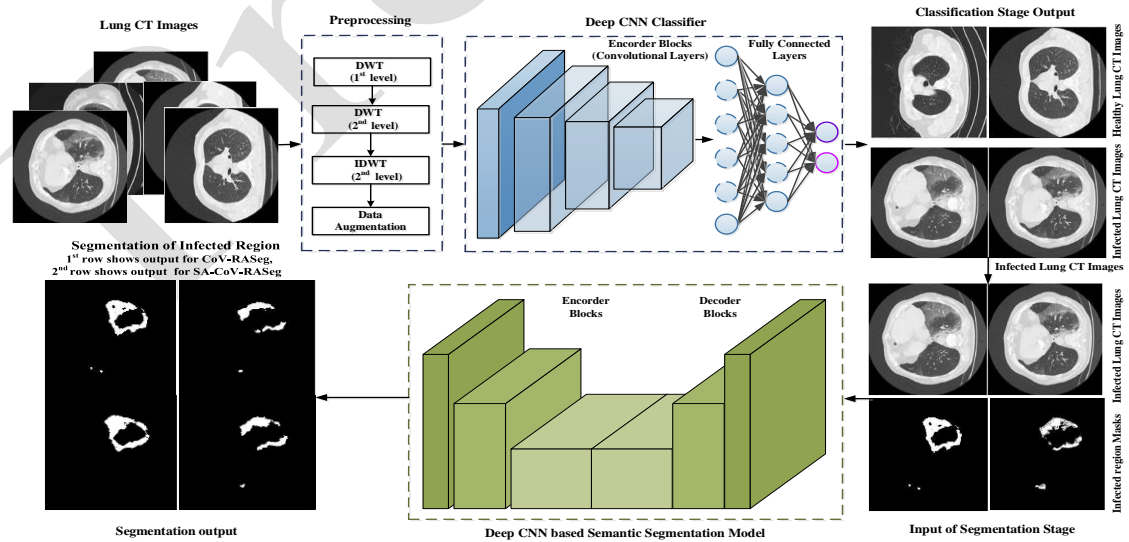

Figure 1: Overview of the workflow for the proposed two-stage deep classification and segmentation framework.

## 3.1. Discrimination of COVID-19 infected samples

In this stage, at a coarse scale, COVID-19 infected CT samples are segregated from healthy samples by performing classification. Initially, a feature space of the dataset is transformed by employing wavelet-based decomposition (shown in Figure 2) to assign class discriminating features to the deep classifiers. In the classification stage, three different experimental setups are implemented: (i) Proposed CoV-CTNet (the detailed explanation of this term will be given in 3.1.2) for classification, (ii) target specific training of deep CNN classifiers from scratch, and (iii) weight transfer-based fine-tuning of deep CNN classifiers. The details of these experimental setups are discussed below.

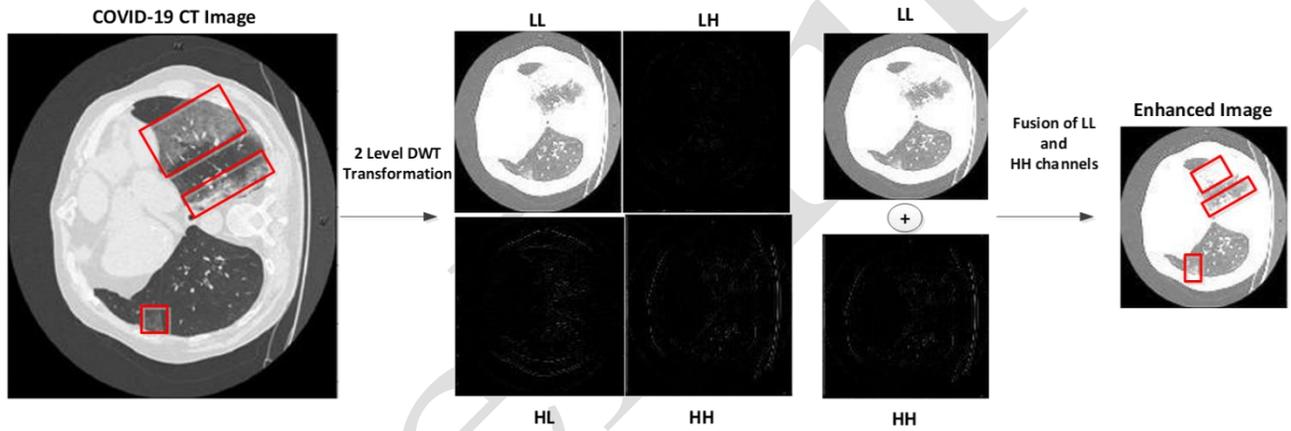

Figure 2: IDWT based Lung CT image enhancement.

### 3.1.1. Feature transformation using wavelet decomposition

The CT image has been subjected to discrete wavelet transformation (DWT) to transform the feature space by decomposing the image into discrete wavelet coefficients using Haar mother wavelet. DWT has two main advantages: (i) reduction in computational complexity, and (ii) image enhancement by transforming the original image into information-rich feature-maps [27], [28].

In DWT, at each decomposition level, the input image is divided into four equal parts: low-low ($D_{i,s=2}^{LL}$), low-high ($D_{i,s=2}^{LH}$), high-low ($D_{i,s=2}^{HL}$), and high-high ($D_{i,s=2}^{HH}$) resolution. Whereas, 'i' represents the level of decomposition (D) and 's' is a scaling factor. In this work, we performed two-level decomposition (shown in Figure 2) to select the highly informative feature-map. For this, the output (LL and HH) from the 1st level decomposition is further processed through DWT transformation to extract information-rich feature-map for classification. The high information features-maps are reconstructed back to images using the Inverse Discrete Wavelet Transform (IDWT). In this study, similar to the idea of Leplacian of Gaussian, we have enhanced the image representation by fusing the images reconstructed from LL and HH channels of second-level DWT representations [29]. This transformed feature space is used to distinguish between healthy and COVID-19-infected images.

### 3.1.2. Proposed CoV-CTNet for classification

In this work, we proposed residual learning-based CNN CoV-CTNet that discriminates COVID-19 infected samples based on CT imagery features. We have used ResNet-18 [30] as a baseline classification model for comparison. The proposed CoV-CTNet is a 24 layers deep architecture consisting of four residual blocks-based feature extraction stages (Figure 3). Within each residual block, two different types of convolutional blocks are implemented (shown in Figure 3). These convolutional blocks are connected with shortcut links within the residual block to perform the reference-based optimization of convolutional filters. A mathematical formulation of the operations used within the convolutional block is expressed in Equation (1-3), whereas the concept of residual learning is expressed in Equation (4) & (5). Residual learning has an advantage over simple feed-forward weight optimization as it solves the vanishing gradient problem, improves the feature-map representation and network convergence.

$$C_{m,n}^l = \sum_{d}^{D} \sum_{m,n}^{M,N} W_{i,j}^l \bullet X_{i+m,j+n}^{l-1} \qquad (1)$$

$$N^l = \frac{C^l - \mu_B}{\sqrt{\sigma_B^2 + \varepsilon}} \qquad (2)$$

$$f(c) = \begin{cases} c & if\ c > 0 \\ 0 & otherwise \end{cases} \qquad (3)$$

$$H^l = Block_n + Block_{n-1} \qquad (4)$$

$$Block_n = H^l - Block_{n-1} \qquad (5)$$

Equation (1) represents convolution operation between image ($X^{l-1}_{i+m,j+n}$) feature-maps and convolutional filter ($W^{l}_{i,j}$) for $l^{th}$ layer, whereas ($M \times N$), and ($D$) represent the spatial dimension and feature-map depth, respectively, for a convolved image. Center coordinates for the convolutional filter are expressed via ($i, j$), and ($C^{l}_{m,n}$) shows the convolved output for ($m, n$) coordinates of the image for $l^{th}$ layer. Equation (2) represents the batch normalization operation for $l^{th}$ layer convolved output ($C^l$), whereas $\mu_B$ and $\sigma_B^2$ represents the mean and variance for a mini-batch. $f(c)$ shows the ReLu activation function and ($c$) is an activation value for each element of feature-map. $Block_n$ and $Block_{n-1}$ show the convolution blocks consisting of convolution operation, batch normalization and ReLu activation for $n$ and $n-1$ layers, whereas Equation (5) shows the residual learning.

The architectural design of the proposed CoV-CTNet is shown in Figure 3. For the regulation of model complexity and learning of invariant features, convolution operation with stride two is performed at the end of each block. In the proposed architecture, max pooling is used on the top of feature extraction stages to retain the most prominent class-specific feature information within the feature-maps.

For classification purpose, two fully connected layers (mathematically express in Equation (6)) are specified to reduce the feature space by globally analyzing their contribution in classification, and in the last, fully connected layer with softmax is used for the discrimination of healthy lung CT samples from infected images. The proposed model is trained on IDWT enhanced images and optimized using cross-entropy loss (represented in Equation (7)).

$$\mathbf{V} = \sum_{k}^{K} \sum_{d}^{D} u_k \bullet C^{l}_{d} \qquad (6)$$

$$-\sum_{n=1}^{2} y_{CT} \log p_{CT} \qquad (7)$$

Equation (6 & 7) represents the fully connected layer, in which $C^{l}_{d}$ is the convolved output having depth $D$, and $u_k$ is $k^{th}$ neuron of $l^{th}$ fully connected layer. Whereas, in the cross-entropy loss (Equation (7)), $p_{CT}$ is the predicted class for input CT image, and $y_{CT}$ is the actual class of the image.

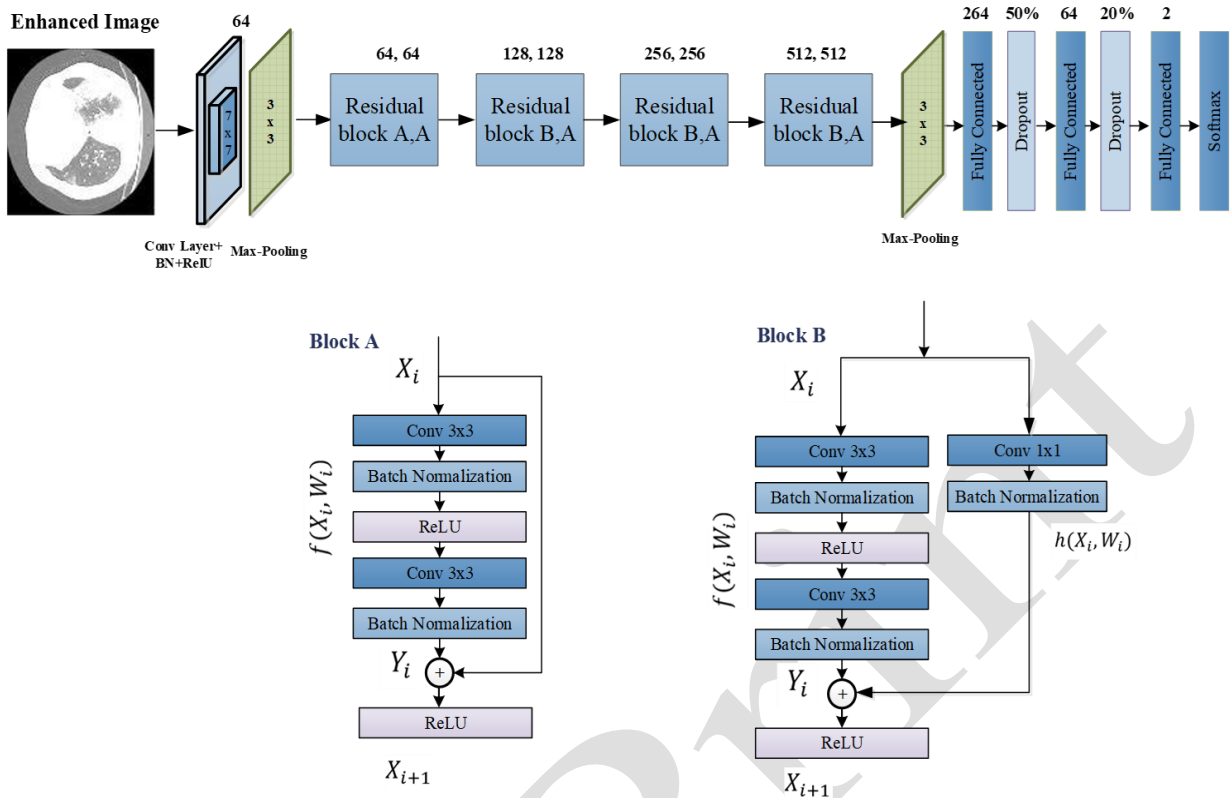

Figure 3: Architectural design for the proposed CoV-CTNet.

### 3.1.3. Performance comparison with existing CNN models

Deep CNNs are a type of DL models that exploit the spatial correlation in images and have shown impressive results for detection, classification, and segmentation related tasks [31]. In recent years, CNN has shown convincing results for biomedical images and has been successfully applied to classify and detect medical images [32]. For comparison of the proposed CoV-CTNet, we applied several existing well-known models varying in depth and architecture design for the classification of COVID-19 infected CT images. The implemented classification networks are VGG16/19, ResNet18/50, GoogleNet (Inception-V1), Inception-V3, DenseNet201, ShuffleNet and Xception [33]–[38]. These networks are trained from scratch (discussed in Section 3.1.3.1) as well as fine-tuned using TL for comparison purpose (discussed in Section 3.1.3.2). Architectural details of analyzed architectures are mentioned below.

VGG is one of the earliest deep architecture that introduced the idea of the effective receptive field by consecutively placing small size filters. It uniformly used 3x3 filters across all the convolutional layers. VGG architecture with two different depths of the feature extraction stage (16 and 19 convolutional layers) and three fully connected layers are proposed for classification. GoogLeNet introduced the idea of inception block, which replaced the conventional convolutional filter within the layer with block architecture. Inception block transforms the image at different scales by using multidimensional filters and concatenating their output. Inception-V3 is a modification of GoogLeNet, which makes the network computationally efficient by replacing the large size filters with asymmetric filters. ResNet proposed the idea of residual learning-based optimization for deep architectures using skips connections. This type of learning improves the network convergence by considering the previous layer output as a reference for the optimization of the next layers' weight. ResNet with various depths depending upon the number of residual blocks has been proposed. DenseNet also exploited the idea of skip connections but in a modified fashion. It concatenates the feature-maps of each subsequent layer with the next layer. In this way, it solves the vanishing gradient problem and provides both high- and low-level features at lateral layers of the network. ShuffleNet is a resource-efficient network, which reduced the number of computations by using point-wise group convolution. Whereby, it retains high accuracy by incorporating the idea of channel shuffle. Xception is a variant of Inception architecture that uses the uniform size of convolutional filters within a block. It increased the number of transformation blocks and gave the idea of depthwise separable convolution.

### 3.1.3.1 Target specific training of deep CNNs from scratch

Seven different well-known existing deep CNN architectures are implemented to evaluate their learning potential for the classification of COVID-19 infected individuals. These deep CNN architectures are made applicable to the targeted CT image dataset by adding a new input layer according to the dimensions of the CT image dataset (82x82x1). Similarly, the last fully connected layer that is used for classification is replaced with a target-specific layer consisting of two neurons. Contrary to this, convolutional blocks from feature extraction stage are kept unchanged in state-of-the-art models. On the last layer, softmax is used to obtain the class-specific probabilities. These models are trained from scratch on CT images, and weight space is optimized using a backpropagation algorithm by minimizing the cross-entropy based loss function.

### 3.1.3.2 Weight transfer-based fine-tuning of deep CNNs

Deep CNN architectures typically demand a sufficient amount of data for effective training. However, currently, the COVID-19 dataset contains a limited amount of publicly available standardized CT images with radiologist's defined ground-truth. The accessibility of labeled data is limited because of patient's privacy concerns and a sufficient amount of time required for pixel-level labelling that is difficult to manage in a time of the pandemic. A small number of samples impede the convergence of deep CNNs while training from scratch [39]. Therefore, in this work, we exploited TL to use the weight vectors of networks pre-trained on a large amount of benchmarked datasets ImageNet. State-of-the-art deep CNN models with optimized filter weights that are learnt from source-domain (ImageNet dataset) are fine-tuned on CT images (target domain) to effectively learn the target-domain specific features from a limited amount of COVID-19 infected patient datasets [40]–[42].

TL is a type of ML, which enables re-utilization of computationally intensive deep CNNs that are already pre-trained on a benchmarked dataset (also called as source domain) having a large number of images for a new problem (known as target-domain) that is comprised of the small training dataset. In TL, the source domain $D(S) = \{X_S, P(X_S)\}$ and target domain $D(Z) = \{X_Z, P(X_Z)\}$ are constitutes of feature space ($X_S$, $X_T$) and its marginal distribution ($P(X_S)$, $P(X_Z)$), whereas their respective tasks ($T(S) = \{Y_S, \eta_S\}$, $T(Z) = \{Y_Z, \eta_Z\}$) are defined by their labels ($Y_S$, $Y_Z$) and objective functions ($\eta_S$, $\eta_Z$). The transfer of knowledge from source to target requires that either both domains have a common task ($\eta_S = \eta_Z$) or they share the feature space ($X_S = X_Z$) or its marginal distribution ($P(X_S) = P(X_Z)$).

There are several types of TL; one of the effective ways of TL is the initialization of parameters of deep architectures from pre-trained models and then fine-tunes the learnable parameters to make them adaptable to the target domain. This strategy is employed when the dataset is small, and the target domain shares a similarity with the source domain in terms of feature space or task [43]. In images, usually, low-level features are common among different categories of images such as curves, edges, gradient, etc., whereas high-level features are class-specific. Based on this assumption, we adapted pre-trained deep architectures for classification of Lung CT images.

In this regard, TL based optimization of existing deep CNN models on target data is performed by adding a new input convolutional layer that coincides with the size of CT samples (i.e., 82x82x1). Similarly, fully connected layers are replaced with the target-domain specific classification layers and dimension of the last layer is aligned to the number of the classes, i.e., two. The model is trained by fine-tuning the learnable layers of feature extraction stage and by optimizing the weights of classification layers from scratch.

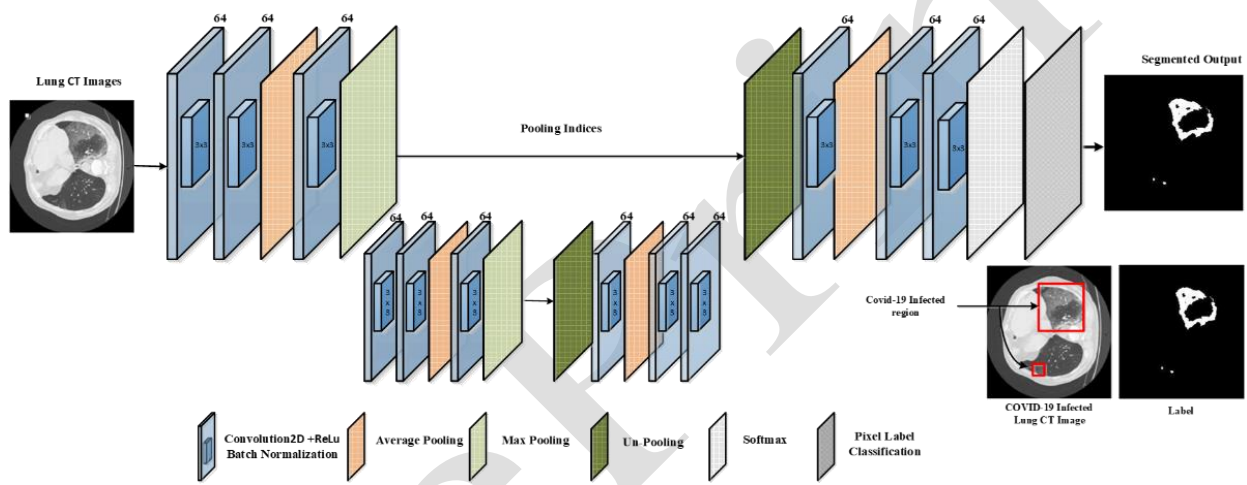

Figure 4: Architectural design for the proposed CoV-RASeg.

## 3.2. Segmentation of COVID-19 infected lung regions

Localization and quantification of the infectious region are crucial for the analysis of infection pattern and its extent in diagnostics. Therefore, after the discrimination of CT images at a coarse-scale, semantic segmentation is performed to obtain subtle inference of infectious regions on CT images. In this work, COVID-19 infected regions are segmented from the surrounding regions by dealing it as a binary semantic segmentation problem. Pixels of the infected regions are labelled as a positive class, whereas all other pixels are regarded as background class.

Semantic segmentation is a fine-scale pixel-based classification that labels each pixel by its corresponding object or region class [44], [45]. In this work, we implemented four different setups for segmentation: (i) proposed COVID-19 based region approximation CNN CoV-RASeg for segmentation, (ii) target specific implementation of deep semantic segmentation models from scratch (iii) pixel attention based implementation of deep semantic segmentation models, and (iv) TL based fine-tuning of deep semantic segmentation models. Details of the experimental setup are mentioned below.

### 3.2.1. Proposed CoV-RASeg for segmentation

The proposed CNN based semantic segmentation architecture CoV-RASeg is like a SegNet, consisting of two encoder and decoder blocks. Architectural design for the proposed CoV-RASeg is shown in Figure 4. In the proposed architecture, we redesign the encoder and decoder block to enhance the network's feature learning capacity. For this purpose, we systematically incorporated average pooling with max-pooling in encoding stages (mathematically expressed in Equation (8 & 9)). In the decoder stage, we also implemented average pooling along with max-un-pooling, in contrast to other deep CNN semantic segmentation models. The difficulty in discrimination of COVID-19 infectious region from a background region is faced as a border between two regions is usually ill-defined, and infectious region overlaps with healthy lungs sections. Max pooling is used to learn the boundary information, whereas average pooling is used to determine characteristic COVID-19 infection patterns from the CT images. We used SegNet as a baseline model to evaluate the significance of systematic using max and average pooling in each encoder and decoder.

$$P_{m,n}^{\max} = f_{\max}(O_{m,n}) \quad (8)$$

$$P_{m,n}^{avg} = f_{avg}(O_{m,n}) \quad (9)$$

In Equation (8 & 9), $f_{avg}(.)$ and $f_{\max}(.)$ represent the max and average pooling operations, respectively, on convolved output ($O_{m,n}$).

We exploited encoder-decoder architecture for fine-grain semantic segmentation as encoding stages of such architectural design are very good in learning of semantically meaningful object-specific information. However, the feature encoding process loses spatial information that is required for object segmentation. Therefore, for the localization of infected regions on the original high-resolution image, we used the decoding stage to nonlinearly restore the spatial resolution of encoder's feature-maps by utilizing max-pooling indices. Whereas, in the last layer, 2x2 convolutional operation with sigmoid activation function is employed for discriminating each pixel into either infected or background region. Encoder and decoder are symmetrical in structure with the difference of max-pooling layer in the encoder part that is replaced by un-pooling layers in the corresponding decoder part (Figure 4).

### 3.2.2. Comparison with existing semantic segmentation models

Several DL models with different architectural designs are reported for semantic segmentation and are benchmarked against diverse categories of the datasets [46]. The differences among these architectures are in the number of encoder and decoders, up-sampling approach, and type of skip connections. In this work, we implemented VGG-16, SegNet, U-SegNet, FCN and Deep LabV1/3 as segmentation models [47]–[50].

To implement the well-known CNN segmentation models on COVID-19 infected lung CT dataset, we replaced the input and classification layers with new layers that correspond with data dimension (304x304x3) and lung mask categories (304x304x2). Initially, deep CNN models are trained from scratch on targeted CT dataset by assigning random values and using a backpropagation algorithm. To overcome the limitation of a small dataset, we also exploited the idea of weight initialization of network layers from pre-trained architectures. In this approach, instead of starting from random weight values, we initialized weights of convolutional layers from pre-trained architectures as discussed in Section 3.1.3.2. The parameters of these models are tuned in an end-to-end manner to adapt diverse categories of models to CT images.

Brief description of the implemented architectures is mentioned in this paragraph to gain an insight into their architectural design. VGG is a state-of-the-art deep architecture, which is famous because of its simple architectural design. VGG based segmentation model is a modified form of classification based VGG architecture in which convolutional layers are used for the encoding of class-specific features, whereas fully connected layers are replaced with a decoder architecture. Decoder architecture is similar to encoder architecture; however; pooling layers are replaced with up-sampling layers. The VGG-16 encoder contains 13, while VGG-19 contains about 16 convolutional layers [38]. FCN performs pixel-to-label based segmentation by using down-sampling and up-sampling stages. The down-sampling stage is used to extract the high-level semantically meaningful information; whereas the up-sampling stage is used to predict the low-level object anatomy (target shape and contour) based information. SegNet is a semantic pixel-wise deep CNN segmentation architecture, inspired by VGG. The key difference between SegNet and VGG is in the number of encoders and decoders. Moreover, SegNet suggested the use of pooling indices in the decoding stages that are stored during the max-pooling based down-sampling in the encoder. The use of pooling indices instead of bilinear interpolation during up-sampling reduces computational complexity. U-SegNet implemented the characteristic attributes of both SegNet and U-Net. U-SegNet leveraged the idea of using skip connections from U-Net architecture for depth-wise concatenation of feature-maps. Whereas, on the decoder side, it uses pooling indices like the SegNet to project the down-sampled feature-maps back to high resolution. Deep Lab implemented Atrous convolution in a cascade manner to regulate the size of feature-map and to view feature-maps at multiple scales.

### 3.2.3. Pixel attention-based implementation of deep CNNs

In this work, we also incorporated the idea of assigning attention to each pixel during training based on their respective class representation in the dataset to deal with the scant representation of COVID-19 infected regions [51]. This is a type of static attention (SA), which enhances the foreground (COVID-19 infected region) by assigning this region high weightage while suppressing the background region pixels by associating small weight with them. It also helps to learn the foreground region anatomies effectively. This idea is incorporated in the proposed CoV-RASeg as well as in the existing deep CNN based semantic segmentation models.

## 4. Experimental setup

### 4.1. Dataset

For the development of the proposed framework, we used standard CT images dataset provided by the SIRM [52]. As COVID-19 is a new disease; therefore, no other standard dataset is available. The dataset is consisting of 829 axial CT samples, out of which 370 CT images show COVID-19 infection pattern, whereas 459 images represent the healthy samples. The provided dataset was examined by the experienced radiologist, and infected lung sections were also marked by the radiologists. Each CT sample was paired with a radiologist provided binary mask (ground truth) that is a fine-grained pixel-level binary label (shown in Figure 5). Moreover, the dataset covers different infection levels like mild, medium, and severe cases of COVID-19. All the images were resized from 512x512x3 to 304x304x3 for computational efficiency. The CT images for COVID-19 infected and healthy lung samples are shown in Figure 5.

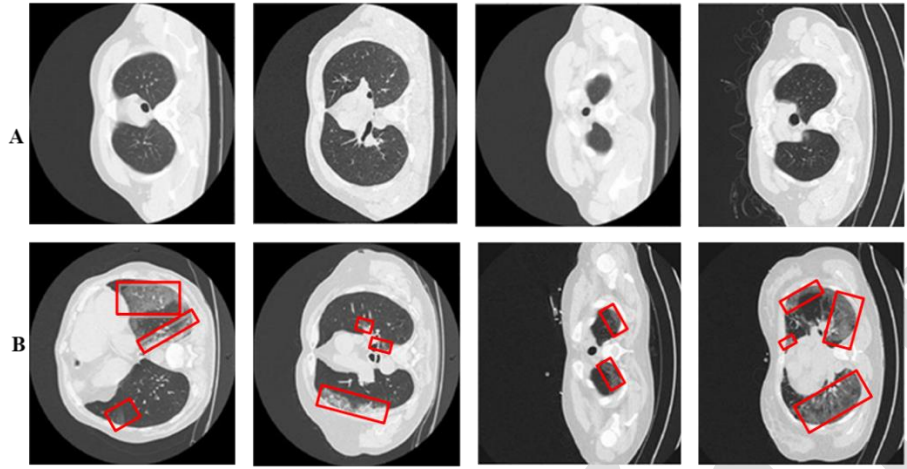

Figure 5: Healthy vs COVID-19 infected Lung CT images are respectively, shown in Panel (A) and (B). Infectious regions are highlighted by red boxes.

## 4.2. Implementation details

In our proposed framework, both the classification and segmentation models are trained separately. For the classification task, the COVID-19 dataset contains 829 training CT images, where 370 are infected images and 459 uninfected images. The 370 affected COVID-19 images, along with their corresponding binary labels, are used during the training of segmentation models. The experimental setup for both classification and segmentation models during training is kept fixed. The dataset is divided into train and test portions at the ratio of 80:20%.

Furthermore, the train set is divided into a train and validation set at a ratio of 80:20% for hypermeter selection. Deep CNNs are trained by employing SGD as an optimizer and by minimizing cross-entropy loss. The hyperparameter values like learning rate: 0.001, Epochs: 30, batch size: 8, and momentum: 0.95 are kept constant. For classification and segmentation, the softmax is used for the identification of class probabilities. 95% confidence interval (CI) for AUC, classification and segmentation models is computed using [53], [54]. All experiments are carried out on MATLAB 2019b framework. Intel (R) Core (TM) i7 and GPU-enabled Nvidia® GTX 1060 Tesla. The training of all the classification and segmentation networks took approximately took ~3 days.

## 4.3. Cross-validation scheme

The cross-validation technique is employed during hypermeter selection to improve the robustness and generalization of the models. We used 5-fold cross-validation during the training of both the classification and segmentation models. Each fold contains 598 training images and 135 validation images in the classification challenge, while the test set includes 166 CT images. Whereas, for the segmentation phase, 370 images are provided with pixel-level masks for COVID-19 infected regions and are involved in pixel-label-based semantic segmentation. On the other hand, 236 images from COVID-19 datasets were considered for training, 60 for validation, and 74 for testing.

## 4.4. Performance evaluation

The proposed two-stage CNN framework performance has been evaluated using standard metrics. Evaluation metrics, along with abbreviation and mathematical explanations, are provided in Table 1. The classification metrics include accuracy (Acc), recall (R), specificity (S), precision (P), Mathew Correlation Coefficient (MCC), and F-score are expressed in Equation (10-14). While the segmentation models are evaluated in terms of segmentation accuracy (S-Acc), the intersection of union (IoU), and the Dice Similarity (DS) coefficient that are expressed in Equation (16) and (17), respectively.

Table 1: Detail of standard classification and segmentation evaluation metrics.

| Metric | Symbol | Explanation |
|---|---|---|
| Accuracy | Acc | Correct predictions |
| Recall | R | The ratio of correct predictions for Covid-19 CT samples |
| Specificity | S | The ratio of correct prediction for Healthy CT samples |
| Precision | P | The similarity of predictions with ground truth |
| True Positive | TP | Correctly predicted Covid-19 CT samples |
| True Negative | TN | Correct predicted Healthy CT samples |
| False Positive | FP | Falsely predicted Covid-19 CT samples |
| False Negative | FN | Falsely predicted Healthy CT samples |
| Positive Samples | TP+FN | Total COVID-19 CT samples |
| Negative Samples | TN+FP | Total Healthy CT samples |
| Jaccard Coefficient | IoU | %Similarity of ground truth and identified regions |
| Dice Similarity | DS Score | %Weighted similarity of ground truth and identified regions |
| Segmentation Acc | S_Acc | %Correctly divided pixels into the positive (Infected) and negative region (Background). |
| True Prediction | TP+TN | Correct classification |
| Negative Prediction | FP+FN | Misclassification |

$$Acc = \frac{Correctly\ Predicted\ Samples}{Total\ Samples} \times 100 \tag{10}$$

$$P = \frac{Correctly\ Predicted\ COVID-19}{Correctly\ Predicted\ COVID-19+ Correctly\ Incorrectly\ Predicted\ COVID-19} \tag{11}$$

$$R = \frac{Correctly\ Predicted\ COVID-19}{Total\ COVID-19\ Samples} \tag{12}$$

$$S = \frac{Correctly\ Predicted\ Healthy}{Total\ Healthy\ Samples} \tag{13}$$

$$F-Score = 2\frac{(P\ x\ R)}{P+R} \tag{14}$$

$$MCC = \frac{(TP\ x\ TN)-(FP\ x\ FN)}{\sqrt{(TP+FP)(FP+FN)(TN+FP)(TN+FN)}} \tag{15}$$

$$IoU = \frac{Correctly\ predicted\ infected\ region}{Correctly\ predicted\ infected\ region+ Total\ infected\ region} \tag{16}$$

$$DS\ Score = \frac{2*Correctly\ Predicted\ infected\ region}{2*Correctly\ Predicted\ infected\ region+Total\ infected\ region} \tag{17}$$

## 5. Results

In this work, we proposed a two-stage framework for the analysis of COVID-19 infected lungs. The advantage of dividing the workflow into two stages is to initially scrutinize the infected CT samples, whereas exploration of regions corresponding to characteristic infection pattern may only be performed within the selected images. This staging process emulates the clinical workflow, where patients based on initial screening are devised for further diagnostic tests. The results of the two stages are discussed below.

### 5.1. Performance analysis of the screening stage

In this work for initial screening, we proposed a deep CNN based classification model CoV-CTNet for categorization of each sample into infected or healthy images. We optimized this stage for highly sensitive in identifying COVID-19 symptoms with a minimum number of false positives. The learning potential of the proposed CoV-CTNet for COVID-19 specific CT feature is evaluated by comparing performance with exiting models on the unseen test dataset.

In the classification stage, we gain improvement in detection rate (Figure 6, Table 2 & Table 3) by adding two enhancements. In the first step, we enhanced the original image by fusing high-frequency channel (HH) with region approximation (LL) channel from the second level decomposition of DWT. This fusion mimics the idea of Leplacian of Gaussian and heightened the distinct characteristics of infected and healthy regions, and thus improves the detection, which is evident from MCC score and other performance metrics (Table 2 & Table 3). Secondly, we proposed a new CNN model CoV-CTNet in which we added fully connected layers with dropout for emphasizing on learning of discriminatory CT based image features for COVID-19 infection.

Table 2: Comparison of the proposed CoV-CT with baseline ResNet-18 on test dataset. 95% CI for AUC is shown in brackets.

| Model | CT sample | %Accuracy | F-score | MCC | AUC [L, H] | Precision | Specificity | Recall |
|---|---|---|---|---|---|---|---|---|
| **Proposed CoV-CTNet** | IDWT | 98.80 | 0.99 | 0.98 | 0.99 [0.97,1] | 0.99 | 0.99 | 0.99 |
| | Original | 97.97 | 0.98 | 0.96 | 0.98 [0.95,1] | 0.96 | 0.96 | 0.99 |
| **ResNet-18** | IDWT | 97.15 | 0.97 | 0.94 | 0.98 [0.95,1] | 0.98 | 0.98 | 0.96 |
| | Original | 97.09 | 0.97 | 0.94 | 0.98 [0.95,1] | 0.95 | 0.97 | 0.96 |

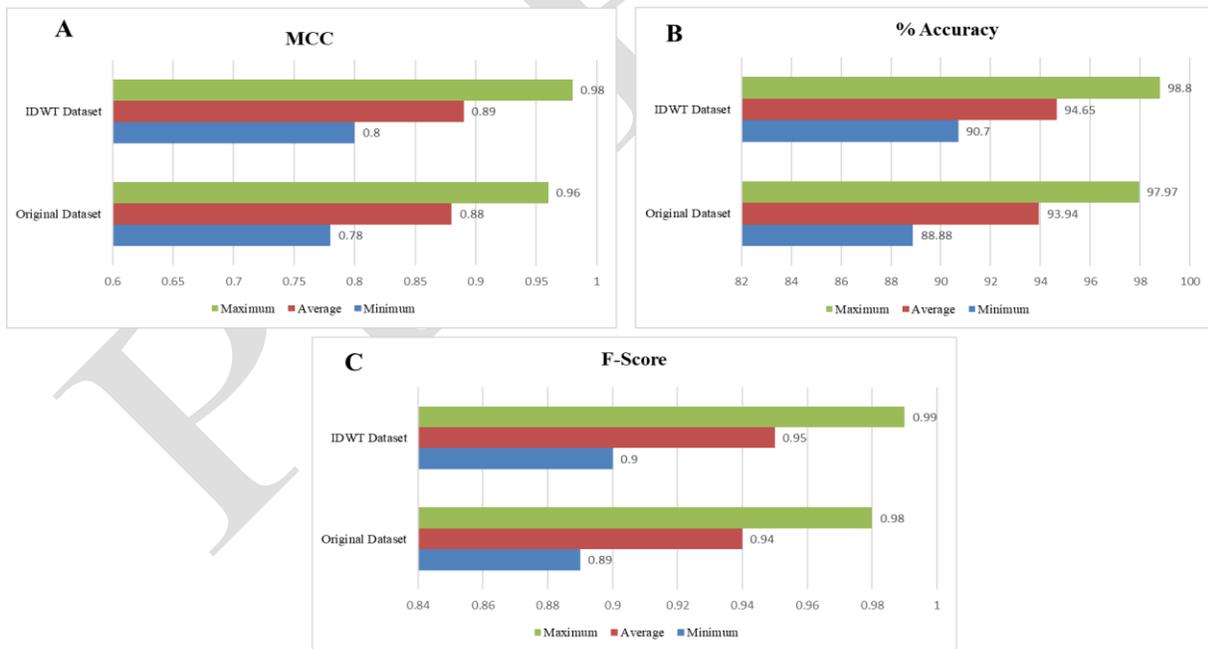

Figure 6: Performance analysis of deep CNN models on IDWT and original images in terms of minimum, average and maximum scores for MCC, Accuracy and F-score.

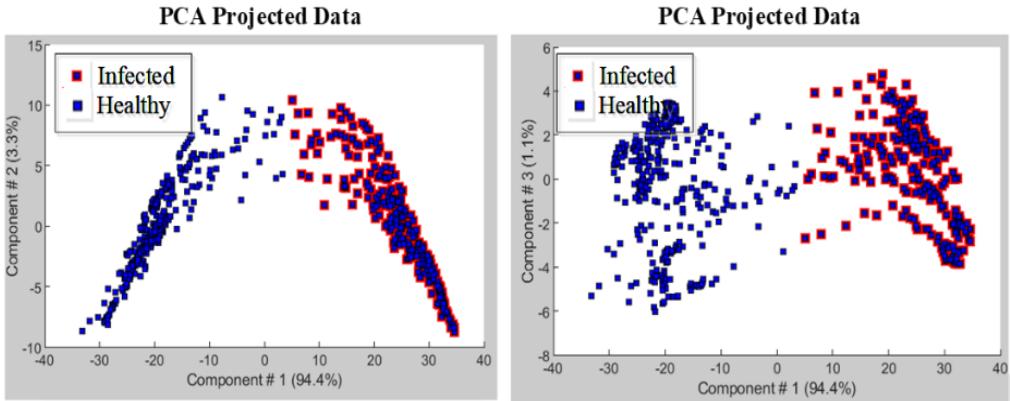

Figure 7: Visualization of feature-space generated from proposed CoV-CTNet using PCA based transformation.

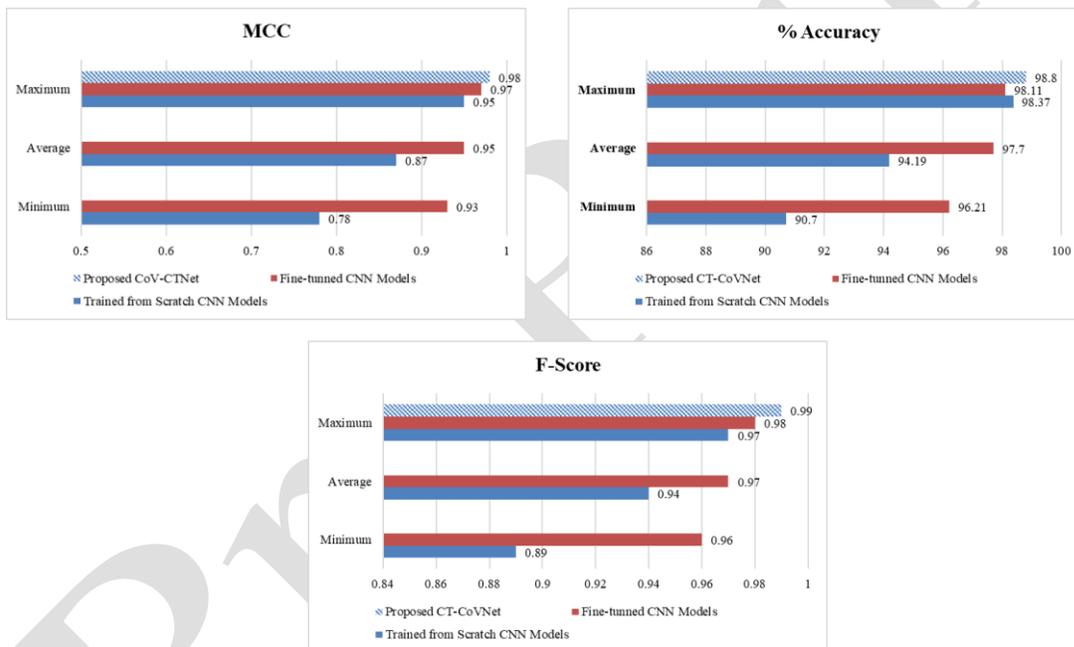

Figure 8: Comparison of the proposed CoV-CTNet with maximum, average and minimum F-scores, MCC and accuracy for state-of-the-art deep CNNs.

### 5.1.1. Performance analysis of the proposed CoV-CTNet

The proposed CoV-CTNet is evaluated on the test set based on three different performance metrics: Accuracy, F-score, and MCC. In addition to this sensitivity, specificity and precision of proposed classifier are also analyzed. The results of the proposed CoV-CTNet are presented in Table 2. The proposed CoV-CTNet shows good generalization as compared to baseline

ResNet18 in terms of F-score (CoV-CTNet: 99%, ResNet-18: 97%), Accuracy (CoV-CTNet: 98.80%, ResNet-18: 97.17%), and MCC: (CoV-CTNet: 98%, ResNet-18: 94%). Moreover, good discrimination ability of the proposed CoV-CTNet is also evident from the decision function feature space (Figure 7).

### 5.1.2. Performance analysis of the proposed CoV-CTNet with existing CNNs

Performance of the proposed CoV-CTNet is compared with the nine different exisiting deep CNN based classification models (VGG-16/19, Inception-V1/V3, ResNet-18/50, DenseNet, Xception, and ShuffleNet). These classification models are well-known for the classification of complex tasks and are successively used for classification of lung abnormalities. For comparison purpose, existing deep CNNs have trained both in a target-specific manner by training from scratch and by fine-tuning them using TL. Table 3 shows the comparison between performances of TL-based fine-tuned and train from scratch models on the test dataset. Performance analysis suggests that TL-based fine-tuned models learn the COVID-19 specific feature in a better way than the deep CNN models that are trained from scratch on CT images.

Whereas comparison of the proposed CoV-CTNet shows better performance in terms of F-score, MCC and accuracy (Table 3) with the existing deep CNNs either they are trained from scratch or fine-tuned. Whereas gain in performance of the proposed CoV-CTNet as compared with best, average and lowest MCC score reported by deep models is shown in Figure 8.

### 5.1.3. PR curve-based analysis

Furthermore, the PR curve is used to quantitatively measure the discrimination power of the proposed model (shown in Figure 9). PR curve is a performance measurement curve that is used for classification problems. It evaluates the generalization ability of the classifier by defining the degree of separability between two classes at different threshold values. Figure 9 shows the obtained PR curve for the proposed and baseline classification models on the test set. It is clearly shown that the proposed CoV-CTNet has a better learning capacity than the baseline and other existing deep CNN models. Although AUC of PR curve for CoV-CTNet is equal to DenseNet, however, overall CoV-CTNet achieved the highest F-score, Accuracy and MCC as compared to DenseNet and other deep models (Table 3).

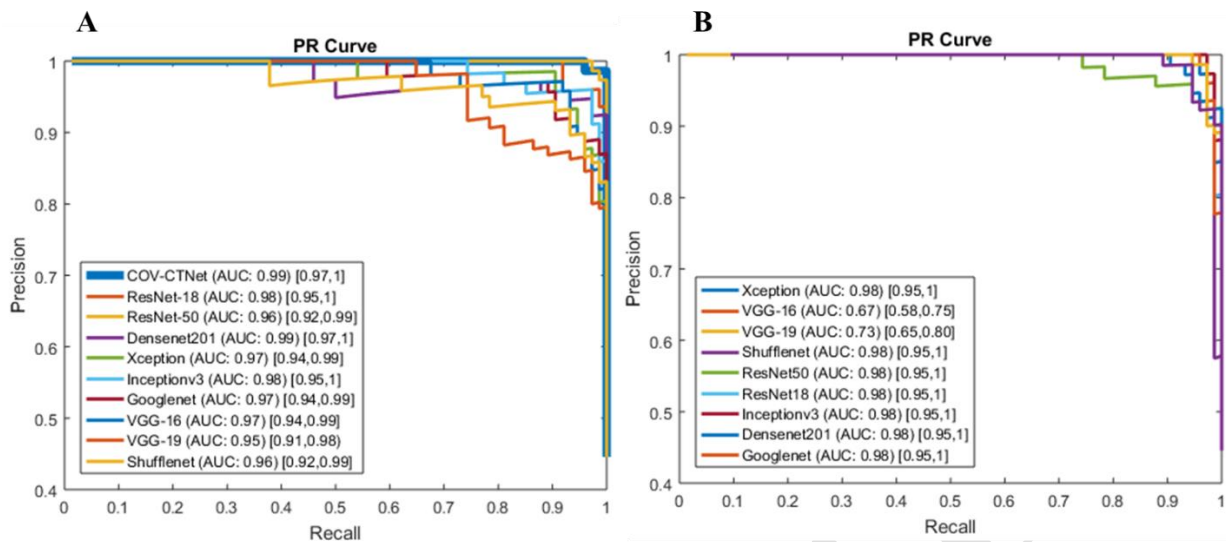

Figure 9: Panel (A) shows the PR curve for the proposed CoV-CTNet and trained from scratch deep CNNs, while (B) shows TL-based fine-tuned existing deep CNNs. 95% CI for AUC is shown in brackets.

Table 3: Comparison between TL-based fine-tuned vs. trained from scratch deep CNN models on the test dataset. 95% CI for AUC is shown in brackets.

| Model | Type | %Accuracy | F-score | MCC | AUC [L, H] | Precision | Specificity | Recall |
|---|---|---|---|---|---|---|---|---|
| VGG16 | Trained from scratch (D) | 91.30 | 0.92 | 0.84 | 0.97 [0.94, 0.99] | 0.98 | 0.98 | 0.85 |
| | Trained from scratch (O) | 91.62 | 0.92 | 0.84 | 0.97 [0.94, 0.99] | 0.96 | 0.96 | 0.92 |
| | Fine-tuned | 97.65 | 0.98 | 0.97 | 0.67 [0.58, 0.75] | 0.97 | 0.97 | 0.98 |
| VGG19 | Trained from scratch (D) | 90.70 | 0.90 | 0.80 | 0.95 [0.91, 0.98] | 0.98 | 0.98 | 0.82 |
| | Trained from scratch (O) | 90.77 | 0.91 | 0.81 | 0.97 [0.94, 0.99] | 0.97 | 0.97 | 0.88 |
| | Fine-tuned | 97.55 | 0.98 | 0.97 | 0.73 [0.65, 0.80] | 0.97 | 0.97 | 0.96 |
| GoogleNet | Trained from scratch (D) | 92.55 | 0.92 | 0.85 | 0.97 [0.94, 0.99] | 0.90 | 0.90 | 0.94 |
| | Trained from scratch (O) | 91.07 | 0.90 | 0.83 | 0.97 [0.94, 0.99] | 0.86 | 0.87 | 0.95 |
| | Fine-tuned | 97.30 | 0.97 | 0.95 | 0.98 [0.95, 1] | 0.94 | 0.94 | 0.98 |
| Inception-V3 | Trained from scratch (D) | 95.52 | 0.96 | 0.91 | 0.98 [0.95, 1] | 0.98 | 0.98 | 0.92 |
| | Trained from scratch (O) | 95.39 | 0.95 | 0.91 | 0.98 [0.95, 1] | 0.97 | 0.97 | 0.93 |
| | Fine-tuned | 96.74 | 0.97 | 0.94 | 0.98 [0.95, 1] | 0.98 | 0.93 | 0.98 |
| Resnet18 | Trained from scratch (D) | 97.15 | 0.97 | 0.94 | 0.98 [0.95, 1] | 0.98 | 0.98 | 0.96 |
| | Trained from scratch (O) | 97.09 | 0.97 | 0.94 | 0.98 [0.95, 1] | 0.95 | 0.97 | 0.96 |
| | Fine-tuned | 98.11 | 0.98 | 0.96 | 0.98 [0.95, 1] | 0.97 | 0.97 | 0.98 |
| Resnet50 | Trained from scratch (D) | 94.45 | 0.94 | 0.89 | 0.96 [0.92, 0.99] | 0.93 | 0.93 | 0.95 |
| | Trained from scratch (O) | 94.75 | 0.95 | 0.90 | 0.98 [0.95, 1] | 0.95 | 0.95 | 0.94 |
| | Fine-tuned | 96.21 | 0.96 | 0.93 | 0.98 [0.95, 1] | 0.94 | 0.94 | 0.97 |
| DenseNet201 | Trained from scratch (D) | 98.37 | 0.98 | 0.97 | 0.99 [0.97, 1] | 0.99 | 0.99 | 0.96 |
| | Trained from scratch (O) | 97.43 | 0.97 | 0.95 | 0.98 [0.95, 1] | 95.95 | 0.96 | 0.99 |
| | Fine-tuned | 98.11 | 0.98 | 0.96 | 0.98 [0.95, 1] | 0.97 | 0.97 | 0.98 |
| ShuffleNet | Trained from scratch (D) | 92.67 | 0.93 | 0.86 | 0.96 [0.92, 0.99] | 0.97 | 0.97 | 0.89 |
| | Trained from scratch (O) | 88.88 | 0.89 | 0.78 | 0.95 [0.91, 0.98] | 0.92 | 0.91 | 0.87 |
| | Fine-tuned | 98.11 | 0.98 | 0.96 | 0.98 [0.95, 1] | 0.97 | 0.97 | 0.98 |
| Xception | Trained from scratch (D) | 94.99 | 0.95 | 0.90 | 0.97 [0.94, 1] | 0.93 | 0.93 | 0.96 |
| | Trained from scratch (O) | 94.45 | 0.94 | 0.89 | 0.98 [0.95, 1] | 0.93 | 0.93 | 0.95 |
| | Fine-tuned | 97.43 | 0.97 | 0.95 | 0.98 [0.95, 1] | 0.95 | 0.96 | 0.98 |

D represents the IDWT images, O represents the Original images

## 5.2. Segmentation of infectious Lung regions

CT images that are detected as a COVID-19 infected in a classification stage using proposed CoV-CTNet are assigned to the deep semantic segmentation model for the exploration of infected regions. The analysis of infected lung lobes is indispensable for gaining radiological an insight into key characteristics of infection pattern, its spread and impact on surrounding lung segments or organs. Moreover, by region analysis, we can quantify infection severity, which may help in the grouping of mildly, and severely infected patients, and their treatment design.

We proposed CoV-RASeg to segment the infected regions of lungs on CT images. Furthermore, a series of deep semantic segmentation models are implemented to assess the learning capacity of our proposed model. The segmentation models were optimized based on three characteristic imagery features of COVID-19 (GGO, pleural effusion and consolidation) to distinguish the typical region on CT image with infected regions. The significance of the proposed method is provided by the experimental results on the test dataset (Table 4).

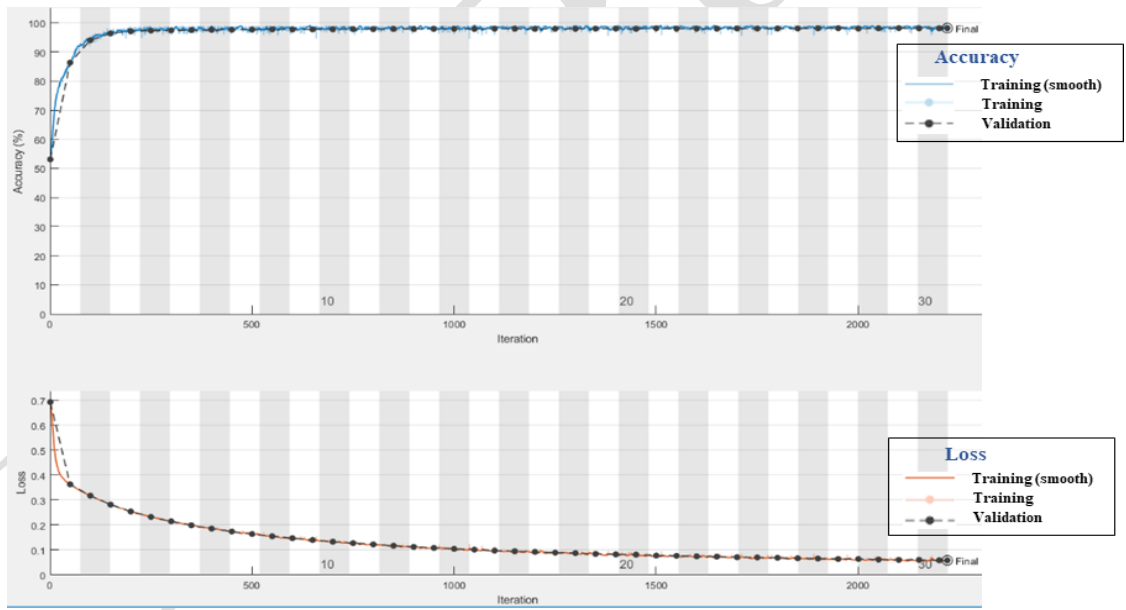

Figure 10: Error convergence plot for the training of the proposed CoV-RASeg segmentation model.

Table 4: Pixel-level classification results for the proposed CoV-RASeg segmentation model.

| Models | Region | % COVID-19 Infected | % Background |
|---|---|---|---|
| CoV-RASeg | COVID-19 Infected (Ground truth) | 99.14 | 0.86 |
|  | Background (Ground truth) | 0.31 | 99.69 |
| SA- CoV-RASeg | COVID-19 Infected (Ground truth) | 99.19 | 0.81 |
|  | Background (Ground truth) | 0.30 | 99.70 |

### 5.2.1. Performance analysis of the proposed CoV-RASeg

Convergences of the proposed CoV-RASeg on train and validation dataset are shown in Figure 10, whereas segmentation capability is evaluated on the test dataset. The main problem in identification of COVID-19 infection is that COVID-19 is characterized by different patterns such as GGO, consolidation, patchy bilateral shadows. Moreover, the pattern of infection varies during the course of information and in different individuals. In the early stages, features of the infected region are indistinguishable from healthy segments. The other important aspect is to isolate the infected region from the healthy region by defining the well-defined boundary.

For delineating the infected regions with well-defined boundary within the lung lobes and dealing with subtle changes, we incorporated the new idea that systematically uses max and average pooling within deep semantic segmentation architecture. The proposed segmentation model suggests good detection ability on the test set, which is evident from DS and IoU score of 95.21% and 98.65%, respectively, for the COVID-19 infected region, as shown in Table 5. Whereas, precise learning of discriminating boundary is clear from the higher value of BFS (97.92%).

The proposed CoV-RASeg is benchmarked against the SegNet. Table 5 illustrates the performance of CoV-RASeg and baseline "SegNet" on the test dataset. DS score and IoU suggest the better performance of proposed architecture as compared to SegNet. Moreover, the pixel-level comparison of the generated binary masks also shows high visual quality for the proposed model compared to baseline (Figure 11). It is apparent from the visual representation that our model is capable of identifying all the infected regions and can highlight their extent precisely. Qualitative analysis (Figure 11) further suggests that our proposed model is useful in segmenting different levels of infection (low, medium, high) in various lobes of the lungs. Moreover, it can localize the infection in a precise manner, whether it is located at a single location or it is spread across multiple distinct lobes or segments of the lungs.

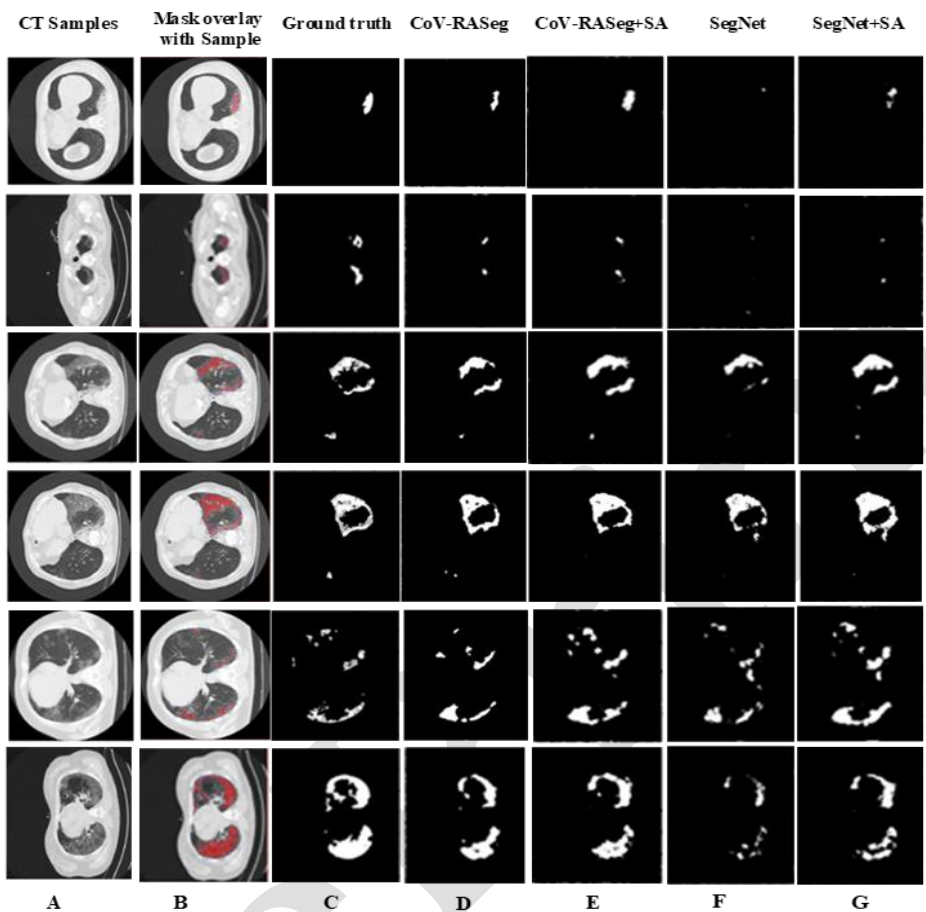

Figure 11: Qualitative assessment of Segmentation results. Panel (A): CT images overlapped with infectious regions that are highlighted with red color, panel (B): binary mask provided by radiologists, panel (C): proposed CoV-RASeg output, panel (D): proposed CoV-RASeg with pixel attention output, panel (E): SegNet output, panel (F): SegNet with pixel attention output.

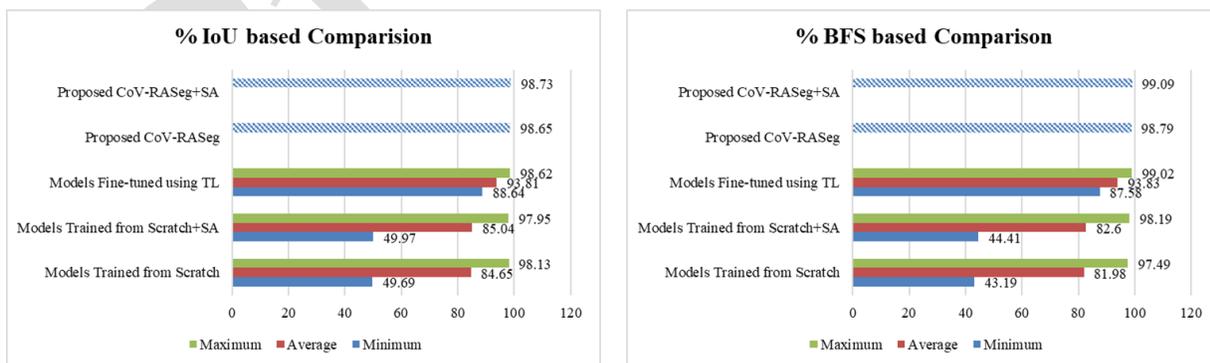

Figure 12: IoU and BFS based comparison of the proposed CoV-RASeg with exiting deep segmentation models.

### 5.2.2. Performance comparison with existing segmentation models

We have validated the performance of our proposed architecture CoV-RASeg by comparing its performance with seven popular semantic segmentation models (VGG16/19, FCN, SegNet, U-Net, U-SegNet, Deep LabV1/3). The comparison is shown in Table 5 & Table 6. The performance of the proposed "RA-CoVSeg" is compared with the existing techniques in three different scenarios, including training from scratch, training using attention and finally, the TL-based fine-tuning of the architecture. The IOU and BFS plot bars show that the proposed model is better or comparable in performance to existing techniques when it is compared with their best, lowest, and average values (shown in Figure 12).

Segmentation output generated by the proposed model CoV-RASeg and state-of-the-art segmentation models are shown in Figure 13 & Figure 14. Qualitative analysis suggests that the proposed CoV-RASeg performs consistently better for all CT samples as compared to other deep semantic segmentation models. It is clear from Figure 13 that existing models perform poorly for the mildly infected CT sample. Whereas, the performance of VGG16, VGG19, FCN and Deep Lab-V1/3 models fluctuate on different categories of CT samples, which suggests the poor generalization. Figure 14 shows that U-Net and a different variant of FCN fail to learn the COVID-19 infection pattern significantly.

For the existing models, TL based Deep LabV3 shows the best performance with DS score: 95%, IoU: 98% and BFS: 99. Whereas, Deep LabV3 fails to learn infection pattern when train from the scratch (Table 5 & Table 6, Figure 13). The worst segmentation performance for COVID infected region is 49.69%, and for the background is 61.06%. Our proposed model, which is small in size shows the comparable performance with high capacity Deep LabV3 fine-tuned models.

Table 5: Performance of deep semantic segmentation models that are trained from scratch on test dataset. Mean error for Dice Similarity (DS) indice is represented at 95% CI.

| Model | Region | DS ± Error | %Acc | %IOU | %BFS | % G_Acc | %M_Acc | %M_IOU | %W_IOU | %M_BFS |
|---|---|---|---|---|---|---|---|---|---|---|
| **SA-CoV-RASeg** | **Infected** | **0.953±0.047** | **99.19** | **98.73** | **99.09** | **99.51** | **99.45** | **98.97** | **99.02** | **98.27** |
| | **Background** | **0.986±0.014** | **99.70** | **99.20** | **97.45** | | | | | |
| CoV-RASeg | Infected | 0.952±0.048 | 99.14 | 98.65 | 98.79 | 99.48 | 99.42 | 98.90 | 98.96 | 97.92 |
| | Background | 0.986±0.014 | 99.69 | 99.15 | 97.05 | | | | | |
| U-SegNet SA | Infected | 0.952±0.048 | 98.63 | 98.13 | 95.85 | 99.28 | 99.23 | 98.48 | 98.56 | 97.15 |
| | Background | 0.979±0.021 | 99.84 | 98.83 | 96.29 | | | | | |
| U-SegNet | Infected | 0.951±0.049 | 98.36 | 97.71 | 97.49 | 99.21 | 99.11 | 98.14 | 98.24 | 96.07 |
| | Background | 0.981±0.019 | 99.86 | 98.57 | 96.82 | | | | | |
| VGG-19 SA | Infected | 0.943±0.057 | 98.97 | 96.56 | 97.73 | 98.72 | 98.71 | 97.18 | 97.32 | 96.48 |
| | Background | 0.979±0.021 | 98.45 | 97.79 | 95.22 | | | | | |
| VGG-19 | Infected | 0.94±0.06 | 98.61 | 95.98 | 96.91 | 98.40 | 98.44 | 96.70 | 96.87 | 95.49 |
| | Background | 0.977±0.023 | 98.28 | 97.42 | 94.07 | | | | | |
| VGG-16 SA | Infected | 0.949±0.051 | 99.09 | 97.95 | 97.99 | 99.20 | 99.18 | 98.33 | 98.41 | 97.13 |
| | Background | 0.984±0.016 | 99.27 | 98.70 | 96.27 | | | | | |
| VGG-16 | Infected | 0.95±0.05 | 98.76 | 98.03 | 97.71 | 99.23 | 99.15 | 98.40 | 98.48 | 97.19 |
| | Background | 0.983±0.017 | 99.53 | 98.76 | 96.66 | | | | | |
| SegNet SA | Infected | 0.949±0.051 | 98.94 | 97.62 | 98.19 | 99.22 | 99.06 | 98.06 | 98.15 | 97.03 |
| | Background | 0.982±0.018 | 99.17 | 98.49 | 95.86 | | | | | |
| SegNet | Infected | 0.951±0.049 | 98.29 | 97.70 | 97.41 | 99.10 | 98.95 | 98.13 | 98.22 | 96.82 |
| | Background | 0.981±0.019 | 99.62 | 98.55 | 96.23 | | | | | |
| FCN-8 SA | Infected | 0.91±0.09 | 91.38 | 88.91 | 89.37 | 95.59 | 94.81 | 91.05 | 91.53 | 83.66 |
| | Background | 0.95±0.05 | 98.25 | 93.18 | 77.95 | | | | | |
| FCN-8 | Infected | 0.907±0.093 | 90.92 | 89.11 | 87.74 | 95.32 | 94.55 | 90.63 | 90.20 | 82.43 |
| | Background | 0.94±0.06 | 98.18 | 92.15 | 77.11 | | | | | |
| Deeplabv3_1 SA | Infected | 0.658±0.342 | 71.88 | 49.97 | 44.41 | 71.53 | 69.33 | 54.40 | 54.14 | 37.04 |
| | Background | 0.696±0.304 | 66.77 | 58.83 | 29.67 | | | | | |
| Deeplabv3_1 | Infected | 0.645±0.355 | 71.83 | 49.69 | 43.19 | 71.88 | 71.87 | 55.38 | 56.67 | 31.97 |
| | Background | 0.745±0.255 | 71.91 | 61.06 | 20.75 | | | | | |
| Deeplabv3_2 SA | Infected | 0.757±0.243 | 79.83 | 66.53 | 54.62 | 82.27 | 80.02 | 70.26 | 72.84 | 39.39 |
| | Background | 0.818±0.182 | 80.21 | 73.98 | 24.16 | | | | | |
| Deeplabv3_2 | Infected | 0.754±0.246 | 80.93 | 63.92 | 53.40 | 82.34 | 82.08 | 69.11 | 70.28 | 39.24 |
| | Background | 0.84±0.16 | 83.23 | 74.29 | 25.08 | | | | | |

SA represents that the models are trained by incorporating static attention

Table 6: Performance of deep semantic segmentation models that are fine-tuned using TL on test dataset. Mean error for Dice Similarity (DS) indice is represented at 95% CI.

| Model | Region | DS ± Error | %Acc | %IOU | %BFS | %G_Acc | %M_Acc | %M_IOU | %W_IOU | %M_BFS |
|---|---|---|---|---|---|---|---|---|---|---|
| SegNet | Infected | 0.946±0.054 | 98.27 | 97.02 | 97.06 | 98.83 | 98.73 | 97.57 | 97.69 | 96.14 |
| | Background | 0.979±0.021 | 99.19 | 98.12 | 95.23 | | | | | |
| VGG16 | Infected | 0.933±0.067 | 97.89 | 94.55 | 95.71 | 97.82 | 97.83 | 95.52 | 95.74 | 93.38 |
| | Background | 0.972±0.028 | 97.77 | 96.49 | 91.05 | | | | | |
| VGG19 | Infected | 0.937±0.063 | 97.95 | 95.37 | 96.03 | 98.16 | 98.12 | 96.20 | 96.39 | 94.53 |
| | Background | 0.974±0.026 | 98.29 | 97.04 | 93.03 | | | | | |
| FCN-8 | Infected | 0.906±0.094 | 91.03 | 88.64 | 87.58 | 95.49 | 94.67 | 90.84 | 91.34 | 83.89 |
| | Background | 0.95±0.05 | 98.30 | 93.04 | 80.20 | | | | | |
| deeplabv3_1 | Infected | 0.951±0.049 | 99.29 | 98.54 | 98.97 | 99.47 | 99.47 | 98.89 | 98.82 | 98.23 |
| | Background | 0.986±0.014 | 99.58 | 99.11 | 97.49 | | | | | |
| deeplabv3_2 | Infected | 0.952±0.048 | 99.48 | 98.62 | 99.02 | 99.54 | 99.53 | 99.03 | 98.88 | 98.42 |
| | Background | 0.987±0.013 | 99.58 | 99.15 | 97.82 | | | | | |

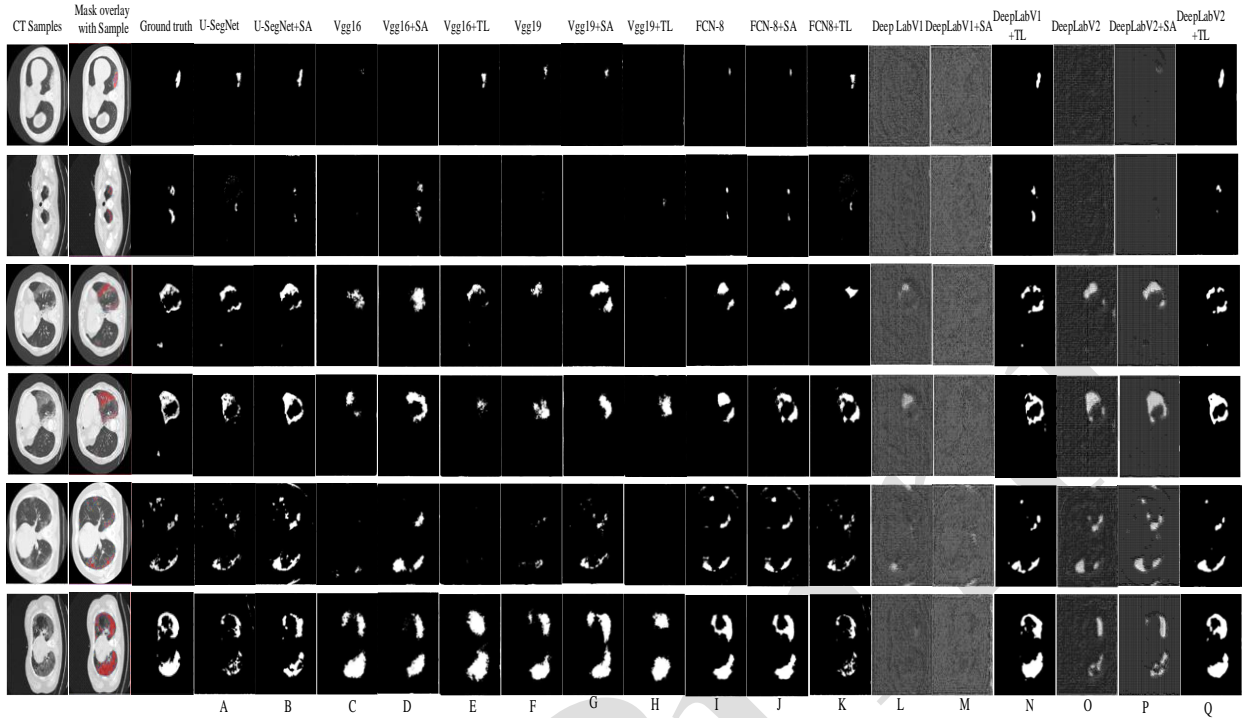

Figure 13: Qualitative assessment of segmentation results on the test set. Panel (A-Q) show the output of different deep segmentation models that are either trained from scratch, fine-tuned using TL or implemented by incorporating pixel attention (SA) during training.

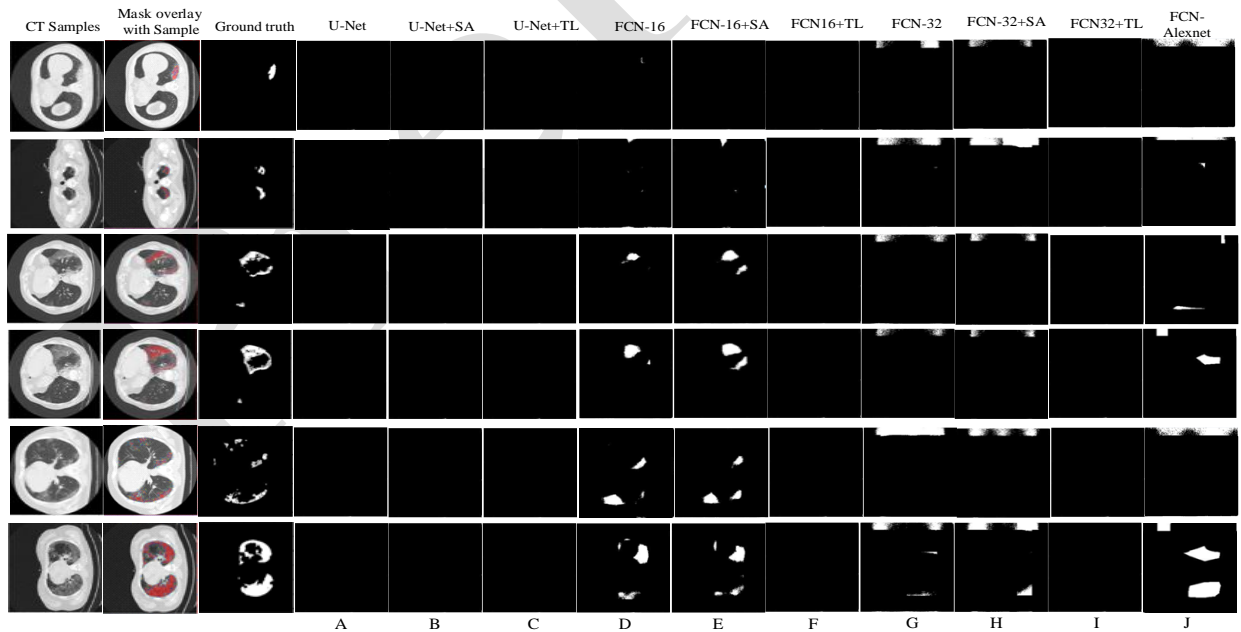

Figure 14: Qualitative assessment of segmentation results for U-Net and Variants of FCN that fail to learn COVID-19 infection pattern for the test set significantly. Panel (A-J) show the output for different deep segmentation models that are either trained from scratch, fine-tuned using TL or implemented by incorporating pixel attention (SA) during training.

### 5.2.3. Performance analysis of attention based deep CNNs

In the provided dataset, the typical region of CT images or healthy samples dominates the COVID-19 infected areas. This under-representation usually affects the performance of segmentation models. To address this problem, we used a pixel attention strategy during training. The incorporation of pixel weights consistently improves the segmentation for different categories of infections, which is evident from the visual quality of the segmentation output maps (Figure 11 & Figure 13) and performance measure (Table 5), whereas significant improvement for less severely infected lung sections is noted. The gain in performance is noted from 0.05 to 0.20%, as shown in Table 5.

## 6. Conclusions

Early detection and analysis of COVID-19 infection pattern is crucial for triage and transmission control. Therefore, in this work, we have proposed a two-stage deep CNN based framework for classification and analysis of COVID-19 infected regions on lung CT images. The good discrimination ability of the proposed CoV-CTNet on test in terms of MCC (0.98) and F-score (0.99) as compared to the existing deep CNNs suggest that it can effectively identify infectious samples with a minimum number of false positives. In this work, we have proposed a region approximation based semantic segmentation model CoV-RASeg for identification and analysis of infected regions on Lung CT images that are classified as COVID-19 using proposed classification model CoV-CTNet. This two-stage processing effectively reduces the search space for learning of characteristic infectious patterns on Lung CT images. The promising performance of the proposed segmentation model (DS: 0.95, BFS: 0.99) suggests that the proposed model can precisely identify the COVID-19 infected regions on lung CT images with subtle details and region boundary. In future, this work can be extended to further classify the infected region into characteristic patterns (GGO, consolidation, pleural effusion) to provide detailed radiological insight into COVID-19 infection pattern.

**Acknowledgment:**
This work was conducted with the support of PIEAS IT endowment fund under the Pakistan Higher Education Commission (HEC). This research was also supported by Basic Science Research Program through the National Research Foundation of Korea (NRF) funded by the Ministry of Education (2014R1A1A2053780).We also thank Pattern Recognition Lab (PR-Lab) and Pakistan Institute of Engineering, and Applied Sciences (PIEAS), for providing necessary computational resources and healthy research environment.